\documentclass[aps,prb,preprint,showpacs]{revtex4-1}

\usepackage{amsmath,epsfig,amsfonts,epsfig,graphicx,
}
\usepackage{color}
\usepackage{graphicx,epstopdf}
\usepackage{color, graphicx}% Include figure files
\usepackage{dcolumn}% Align table columns on decimal point
\usepackage{bm}% bold math
\usepackage{amssymb}
\usepackage{amsmath}
\usepackage{latexsym}
\usepackage{amsfonts}
\usepackage{amsmath}
\usepackage{multirow}
\usepackage{ifthen}

\begin{document}

\title{Slow sound propagation in lossy locally resonant periodic structures}

\author{G. Theocharis}
\author{O. Richoux}
\author{V. Romero-Garc\'ia}
\author{V. Tournat}
%\\
\affiliation{LUNAM Universit\'e, Universit\'e du Maine, CNRS, LAUM UMR 6613, Av. O. Messiaen, 72085 Le Mans, France}

\begin{abstract} We investigate the sound propagation in an air-filled tube periodically loaded with Helmholtz resonators. By tuning the Helmholtz with the Bragg resonance, we study the efficiency of slow sound propagation in the presence of the intrinsic viscothermal losses of the system.  While in the lossless case the overlapping of the resonances results in slow sound induced transparency of a narrow frequency band surrounded by a strong and broadband gap, the inclusion of the unavoidable losses imposes limits to the slowdown factor and the maximum transmission. Experiments, theory and finite element simulations have been used for the characterization of acoustic wave propagation. Experiments, in good agreement with the lossy theory, reveal the possibility of slowing sound at low frequencies by 20 times. A trade-off among the relevant parameters (delay time, maximum transmission, bandwidth) as a function of the tuning between Bragg and Helmholtz resonance frequency is also presented. 
\end{abstract}


\maketitle

\section{Introduction}
Locally resonant acoustic metamaterials \cite{Lu} derive their unique properties 
e.g. negative effective mass density \cite{Ping} and negative bulk modulus \cite{Fang},
from local resonators contained within each unit cell of engineered structures. 
Due to these effective parameters, a plethora of fascinating phenomena have been proposed over
the last years, including negative refraction, super-absorbing sound materials, acoustic focusing, and cloaking 
(see Ref. [\onlinecite{books}] and references therein). 

Although the inclusion of losses in locally resonant structures is very important, their role has been underestimated while in some studies totally ignored. Loss is not only an unavoidable feature, but also it may have deleterious consequences on some of the novel features of metamaterials\cite{meta} including double negativity and cloaking. Recent works on both photonic\cite{Huang04, Pedersen, McPhedran} and phononic\cite{Moiseyenko11, Hussein09, Andreassen13,Laude} periodic structures show that the dispersion relation can be dramatically altered.
%It is found that the group velocity curves, which are fundamental for the study of the propagation dynamics in periodic media, are dramatically altered as losses are considered\cite{McPhedran, Hussein09}.
In particular, flat propagating bands corresponding to slow-wave propagation, acquire an enhanced damping as compared to bands with larger group velocities\cite{Pedersen, Moiseyenko11}. 
%Therefore, losses particularly impact the slow-wave applications\cite{Reza, Tsakmakidis}. 

The aim of this work is to study the influence of losses on slow sound propagation in periodic locally resonant structures. 
For this reason, we theoretically and experimentally analyze the sound propagation in a tube periodically loaded with Helmholtz resonators (HRs) taking into account the viscothermal losses\cite{Zwikker}. In particular, we investigate configurations where the Bragg resonance frequency due to periodicity and the frequency of the Helmholtz resonators either coincide or are very close to each others. 
%In this set-up the HRs are not directly coupled each other, but are coupled through the waveguide. 
In the first case, 
%namely once the Bragg resonance frequency due to periodicity and the frequency of the Helmholtz resonators coincide, 
a super-wide and strongly attenuating band gap is created. This property has recovered interest during last years in different branches of science including elastic waves\cite{Xiao}, split-ring microwave propagation \cite{Paris}, sonic crystals \cite{Page}, and duct acoustics \cite{Wang}, among others. In acoustics, this tuning, first studied by Sugimoto\cite{Sugimoto}, is of great importance for sound and vibration isolation\cite{Wang,Seo}. In the case of slightly  detuned resonances, i.e. once the Bragg and the local resonance are slightly different, an almost flat band appears, a feature which is particular useful for slow waves applications\cite{Boudouti}. Here, we make use of this detuning to theoretically and experimentally examine the effect of losses in the slow sound band. We focus on both periodic systems and finite periodic arrays with $N$ side HRs. %(which has the particularity of presenting $N-1$ flat bands in the lossless case). 
Using the transmission matrix method, we characterize the group index, the bandwidth, and the slow-wave limits of these structures, showing good agreement with experiments. The limit of the slow sound due to losses is of relevant importance for the design of narrow-band transmission filters and switches. Moreover, it could also open perspectives in the way to control the nonlinear effects at the local resonances \cite{PRERichoux}, which could increase the functionality of the acoustic metamaterials leading to novel acoustic devices for the sound control at low frequencies. 

\section{Theory}
The propagation of linear, time-harmonic acoustic waves in a waveguide periodically loaded by side branches has been first studied in Ref. [\onlinecite{Bradley}]. Using Bloch theory and the transfer matrix method, one can derive the following dispersion relation (see also Refs. [\onlinecite{Sugimoto}],[\onlinecite{EPLOlivier}]):

\begin{equation}
\label{eq2}
  \cos (qL) = \cos(kL)+ j\frac{Z_{0}}{2Z_{b}}\sin(kL),
\end{equation}
where $q$ is the Bloch wave number, $k$ is the wave number in air, $L$ the lattice constant, $Z_{b}$ the input impedance of the branch (see Ref. [\onlinecite{EPLOlivier}] for the case of HR branch), and $Z_{0}=\rho_{0}c_{0}/S$ the acoustic impedance of the waveguide where $S$ is its cross-sectional area; $\rho_{0}, c_{0}$ the density and the speed of sound in the air respectively, and $j=\sqrt{-1}$.   

The transmission coefficient through a finite lattice can be derived using the transmission matrix method. For the case of $N$ side branches, the total transmission matrix can be expressed as follows \cite{Pierce,Seo}

\begin{eqnarray}
\label{eq1}
 \left( \begin{array}{c} P_{1} \\ U_{1} \end{array} \right) & = &
(M_b M_T)^{N-1}M_b \left( \begin{array}{c} P_2 \\ U_2 \end{array} \right) , \nonumber \\ 
  & = &  \left( \begin{array}{cc}
  T_{11} & T_{12} \\
  T_{21} & T_{22} 
  \end{array} \right)\left( \begin{array}{c} P_2 \\ U_2 \end{array} \right),
\end{eqnarray}
where
\begin{eqnarray}
\label{eq2}
M_T &=&  
 \left( \begin{array}{cc}
  \cos(kL) & jZ_0 \sin(kL) \\
  \frac{j}{Z_0} \sin(kL) & \cos(kL) 
  \end{array} \right),\\ 
M_b &= &
 \left( \begin{array}{cc}
  1 & 0 \\
  1/Z_b & 1 
  \end{array} \right) ,
\end{eqnarray} 
represent the transmission matrices for the propagation through a length $L$ in the waveguide and through a resonant branch respectively. $P_1$ ($U_1$) and $P_2$ ($U_2$) are the pressure (and respectively volume velocity) at the entrance and at the end of the system. Considering the previous equations, the pressure complex transmission coefficient can then be calculated \cite{Seo} as

\begin{equation}
\label{Trans}
  t =\frac{2}{T_{11}+T_{21}/Z_0+T_{21}Z_0+T_{22}}.
\end{equation}

The sound waves are always subjected to viscothermal losses on the wall and to radiation losses. Viscothermal losses are taken into account by considering a complex expression for the wave number. In our case, we used the model of losses from Ref. [\onlinecite{Zwikker}], namely we replace the wave number and the impedances by the following expressions
\begin{eqnarray}
k=\frac{\omega}{c_0}(1+\frac{\beta}{s}(1+(\gamma -1)/\chi))\\
Z_c = \rho_0 c_0( 1+\frac{\beta}{s}(1-(\gamma -1)/\chi))
\end{eqnarray} 
by setting $s=r/\delta$ where $\delta=\sqrt{\frac{2\mu}{\rho_0\omega}}$ is the viscous boundary layer thickness, $\mu$ being the viscosity of air, $\chi=\sqrt{P_r}$ with $P_r$ the Prandtl number, $\beta=(1-j)/\sqrt{2}$, $\gamma$ the heat capacity ratio of air and $r$ the radius of the considered tube. 
%As one can see, the higher the frequency of the propagating waves and the smaller the radius of the tubes, the higher %the losses. 
Radiation losses, which appear at each connection between the waveguide and the HRs, are accounted for through a length correction of the HRs neck defined in the description of the experimental set-up.

\section{Experimental apparatus}

\begin{figure}
\includegraphics[width=8cm]{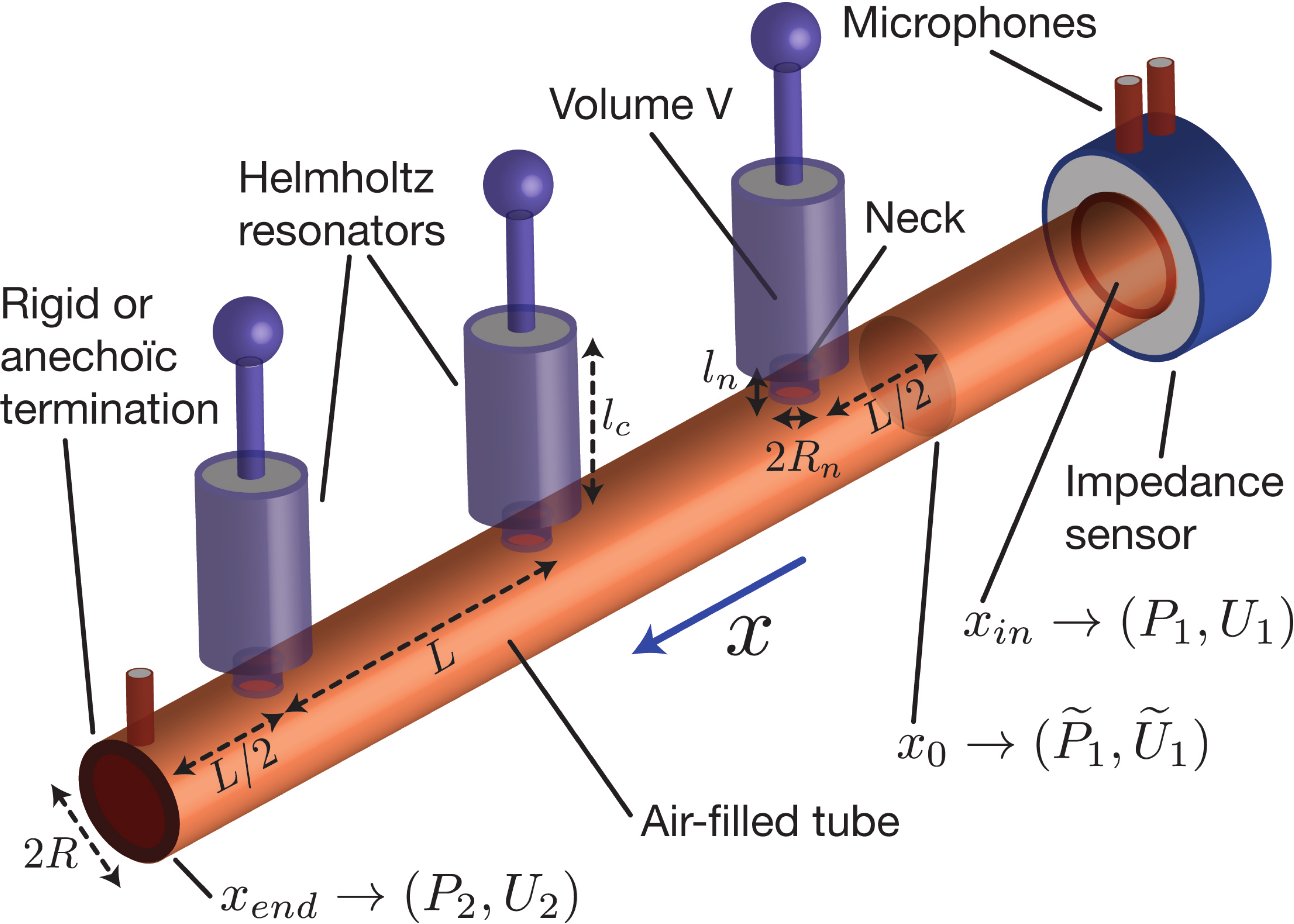}
\caption{(Color online) Schematic of the experimental apparatus.}
\label{Schem}
\end{figure}

The experimental apparatus that we used in this work to calculate the dispersion relation of periodic systems and the transmission coefficient of a finite periodic locally resonant system is shown in Fig. \ref{Schem}. Each HR is made of a neck (cylindrical tube with an inner radius $R_{n}=1$ cm and a length $l_n=2$ cm), and a cavity (cylindrical tube with an inner radius $R_{c}=2.15$ cm and a variable length, $l_c$). We use different configurations through this work from $3$ to $6$ HRs, loaded periodically along a cylindrical waveguide with an inner radius $R=2.5$ cm, $0.5$ cm wall thickness, and total length of $3$ m. The last HR is always connected at a distance of $L/2$ cm from the end of the set-up, $x_{end}$, where $L=30$ cm is the constant distance between adjacent resonators. The sound source is a piezo-electric buzzer embedded in the impedance sensor \cite{ImpedanceSensor} which is placed at $x_{in}$. One BK 4136 microphone, carefully calibrated, is placed at the other end (rigid termination) of the cylindrical waveguide. The frequency range of the applied signal is below the first cutoff frequency of the waveguide, $f_{c01}=4061$ Hz, and thus the propagation can be considered one-dimensional. The end correction of the neck was experimentally measured to be $1$ cm by comparing the input impedances (experimentally and theoretically) for different volumes of the HR cavity. For that purpose, one HR connected to a tube of a length of $15$ cm, rigidly closed, has been used.

\begin{figure*}
\includegraphics[width=15cm]{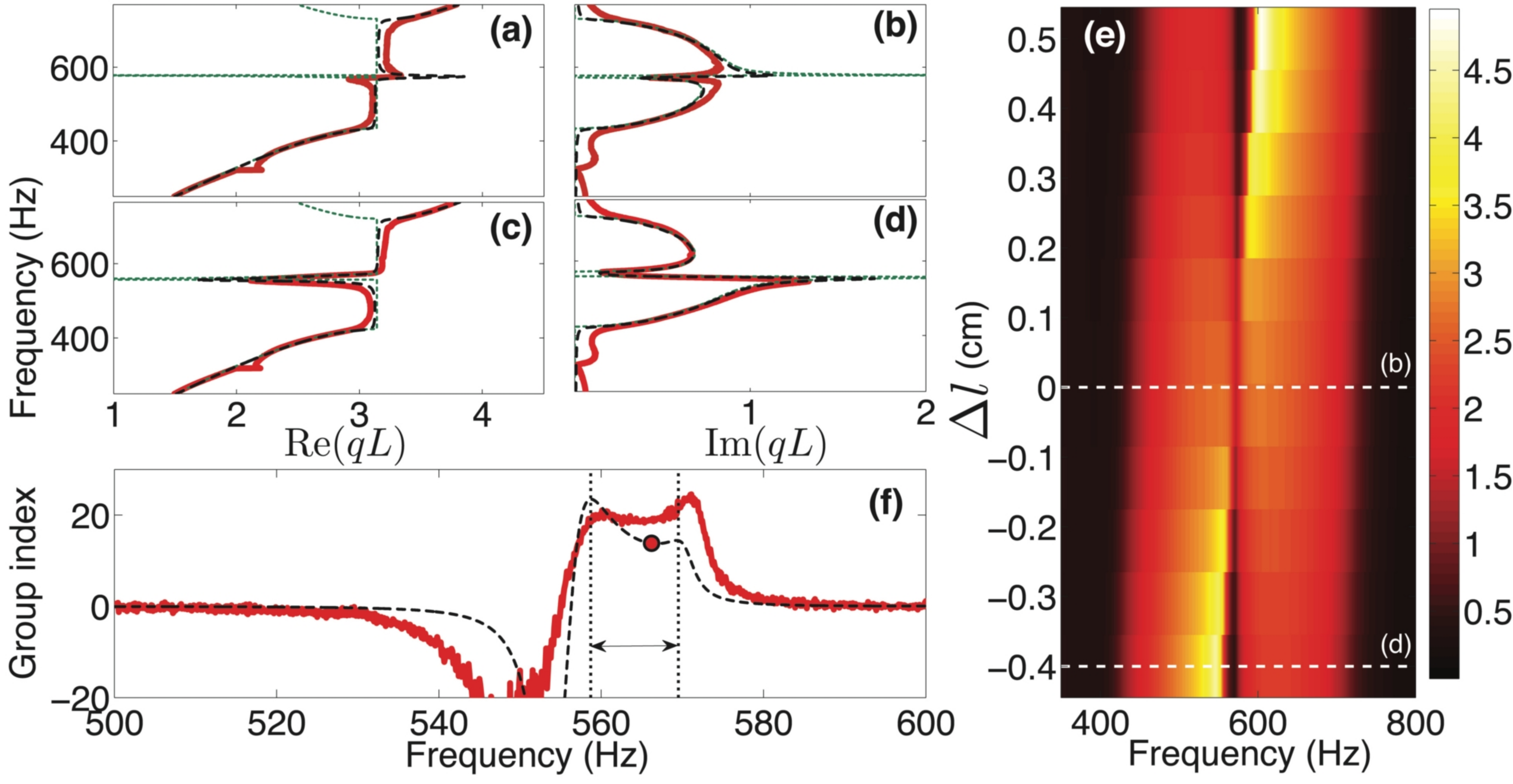}
\caption{(Color online) (a-d) Representation of the complex dispersion relation for the case of $\Delta l=0$ cm (a)-(b), and $\Delta l=-0.4$ cm (c)-(d). Black dashed line shows the analytical lossy case. Green dotted line represents the analytical lossless case. Red solid line shows the experimental result. 
%and red solid lines show the results for the lossy case and green dotted line represents the lossless case. 
(e) Experimental values of Im($qL$) as a function of the length of the cavity. (f) Experimental (red continuous line) and theoretical (black dashed line for lossy case) group index with $\Delta l=-0.4$ cm.}
\label{DR}
\end{figure*}

The input impedance measurement setup (see Fig. \ref{Schem}), together with the transmission matrix method allow us to experimentally evaluate the dispersion relation of periodic systems. To do that we use the measured input impedance $Z=\frac{P_1}{U_1}$ and the transfer impedance $Z_{T}=\frac{P_2}{U_1}$. From $Z$, one can calculate the acoustic impedance at the position $x_{0}$, $\widetilde{Z}=\frac{\widetilde{P}_1}{\widetilde{U}_1}$, which is located $L/2=15$ cm from the first HR as well as the $\widetilde{Z}_{T}=\frac{P_2}{\widetilde{U}_1}$ (see Fig. \ref{Schem}). Then, the impedance matrix of the symmetric structure from $x_{0}$ until the rigid end is given by:
\begin{eqnarray}
\label{Zmatrix}
  \left( \begin{array}{cc}
  \widetilde{Z} & \widetilde{Z}_{T}  \\
  \widetilde{Z}_{T} & \widetilde{Z} 
  \end{array} \right), 
\end{eqnarray}  
from which the transmission matrix of the symmetric system is deduced. Once, we have the transmission matrix of the symmetric periodic structure, we can calculate the Bloch wave number $q$, using $q = \frac{1}{NL}\arccos(\frac{\widetilde{Z}}{\widetilde{Z}_{T}})$\cite{Lissek,Caloz} and as a consequence the dispersion relation of the system. The exact shape of the dispersion curve is obtained by phase unwrapping and by restoring the phase origin \cite{Caloz}. Due to the presence of imperfect matching at the end caused by the rigid termination, some finite size effects are expected in the experimental characterization.

This setup can be also used for the experimental calculation of the transmission coefficient. To do that we replace the rigid end termination by an anechoic termination made of a $10$ m long waveguide partially filled with porous plastic foam to suppress as much as possible the back propagative waves. In this case, the microphone is placed at a distance of L=$30$ cm after the last HR, namely the same distance as between the source and the first HR. Using the transmission matrix method, considering anechoic termination, and assuming the symmetry of the structure, the complex transmission coefficient $t$ reads as follows

\begin{equation}
t=\frac{2 Z_T}{Z+Z_0}.
\end{equation}

\section{Results and discussion}

We start by studying the coupling between the Bragg and the resonance band gap taking into account the presence of the viscothermal losses. In Fig.~\ref{DR}(a)-\ref{DR}(d), one can see both the experimental (red continuous line) and theoretical (black dashed line) complex dispersion relations for two different resonant frequencies of the HRs in comparison with the theoretical lossless case (green dotted line). The experimental setup is composed of $3$ HRs. The slight differences between theory and experiments are due to finite-size effects (see for example Fig.~\ref{DR}(a)-(d) at $f\simeq323$ Hz). 
%and possible inaccuracies in the experimental calculation of the effective neck length. 
For the fixed lattice distance of $L=30$ cm, the first Bragg resonance appears at $f_{B}=\frac{c_0}{2L}\approx 571$ Hz.  The resonance frequency of the HRs $f_{0}$ is in general unknown.  One can use the traditional lumped-parameter model \cite{Pierce} to obtain an analytical expression. However, this is valid only at very small frequencies and it requires the knowledge of the end corrections. Therefore, in order to tune the resonant frequency with the Bragg's frequency, we experimentally calculate the dependence of the imaginary part of the complex dispersion relation on the length of the cavity $l_c$ of the HR (see Fig.~\ref{DR}(e)). We define a detuning length parameter, $\Delta l=l_0-l_c$, where $l_0$ corresponds to the cavity length at which $f_0=f_B$. Thus, $\Delta l$ measures how far we are from the complete overlap between the Bragg and the HR resonance. As shown in Fig.~\ref{DR}(e), if $\Delta l<0$ ($\Delta l>0$) the HR resonance approaches the Bragg's one from lower (higher) frequencies.  According to Ref.~[\onlinecite{Sugimoto}], for the case of $\Delta l=0$ a wide band-gap appears in the region $f_{B}(1-(\kappa/2)^{1/2})<f<f_{B}(1+(\kappa/2)^{1/2})$, where $\kappa=\frac{S_c l_c}{SL}$ measures the smallness of the cavity's volume relative to the unit-cell's volume. 
For our case, white dashed line in Fig. \ref{DR}(e), the above expression predicts a band gap for $416.5<f<727.7$ Hz, which is in very good agreement with the experiments (Fig.~\ref{DR}(a)-(b)) for the case of $\Delta l\simeq0$ cm. However, it is very difficult to find in practice the case of $\Delta l=0$ because one needs to control either the length of the cavity or the lattice constant with a high precision. When $\Delta l \approx0$ one can observe that the lossless theory (green dotted line in Fig. \ref{DR}(a)) predicts a flat branch inside the band gap. This branch is drastically reduced once losses are introduced. 
%On the other hand, from Fig.~\ref{DR}(e), one can notice the increase of the bandwidth of the Bragg band gap due to its %interaction with the Helmholtz band gap at the overlap ($\Delta\simeq0$ cm). As it was pointed out in %Ref.~[\onlinecite{Paris}], this is explained by the increase of the scattering cross section of the resonators as we %approach its resonance frequency.

%{\it Slow-sound transparency and trade-offs of delay and transmission losses for a finite lattice of HRs}: 

We continue by studying the detuned case, i.e., the case $\Delta l\neq0$. In this situation, as shown in Fig.~\ref{DR}(e) between the Bragg and the resonance bands, there is a range of frequencies with small attenuation (small Im($qL$)). For example the real part of the complex dispersion relation for the case $\Delta l=-0.4$ cm (see Fig.~\ref{DR}(c)), shows an almost flat real band, which means slow sound propagation. We introduce the group index as a slowdown factor from the speed of sound $c$, defined as $n_g\equiv c/\upsilon_{g}$ where $\upsilon_{g}= \left(\frac{\partial \omega}{\partial Re(q)}\right)$ is the group velocity. In lossy periodic structures, the real and imaginary parts of the group velocity correspond to propagation velocity and pulse reshaping respectively (see \cite{McPhedran} and references within). Negative values of $n_g$ correspond to negative group velocity induced by the losses, as it has been also reported in Ref.~[\onlinecite{Fang}]. Figure \ref{DR}(f) shows the theoretical and experimental group index obtained from the data shown in Fig.~\ref{DR}(c). It is worth noting that through the bandwidth of the transmitted frequencies (marked in Fig.~\ref{DR}(f) with the double arrow) the group index is $n_g>20$. This slowdown factor is comparable with the results of previously reported experiments \cite{slowsound}.
%, so losses dramatically change the slow down character of the waves inside of this range. 
Figure \ref{DR}(e) shows also that the bandwidth of the transmission band depends on $\Delta l$. Now, we investigate in more detail this slow sound propagation in lossy finite lattice of HRs. In particular, the slow sound propagation is characterized using a delay time and we study the trade-offs among this delay and the transmission losses in a finite lattice.

\begin{figure}
\includegraphics[width=8.5cm]{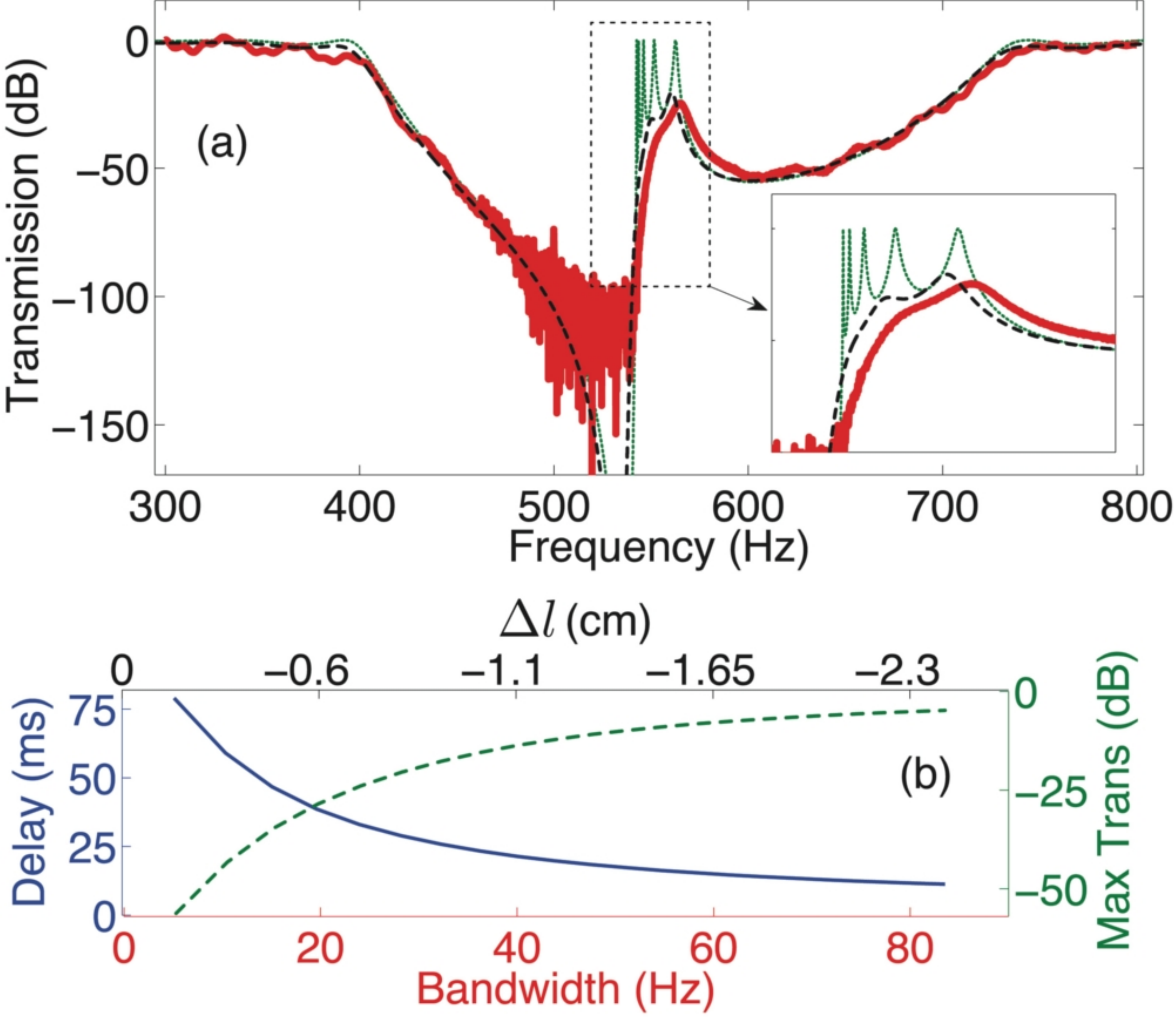}
\caption{(Color online)(a) Transmission coefficient vs. frequency for a lattice of 6 HRs with $L=30$ cm and $\Delta l=-0.8$ cm. Experimental data (red solid line) and their comparison with the lossless (black dashed line) lossy theory (green dotted line). (b) Trade-offs among delay (solid blue) and maximum transmission (dashed green) for a lattice of 6 HRs as a function of the bandwidth and $\Delta l$.}
\label{TGT}
\end{figure}  

Figure \ref{TGT}(a) shows the experimental transmission amplitude in comparison with both the lossless and the lossy theoretical predictions for a lattice of $N=6$ HRs for the detuned case of $\Delta l=-0.8$ cm. 
%We change here to the case $\Delta=-0.8$ cm because as we will see later, once the losses are introduced, the transmitted amplitude depends on the detuning length. For the case of $\Delta l=-0.4$ measurements are practically impossible. 
For the lossless case, there is not a wide transmission band, but $N-1$ narrow transparent ($T=|t|^2=0$ dB) peaks as shown in the inset of Fig.~\ref{TGT}(a). Comparing the lossy with the lossless case, one can observe that the losses reduce and smooth the transmission amplitude of the $N-1$ peaks creating a broadband of transmitted frequencies in good agreement with the experiments. Thus, losses reduce the transparency but create a bandwidth of transmitted modes with similar values of transmission.

In order to analyze the slow down of sound in this transmitted range of frequencies, we define the time delay of a pulse propagating through the whole length $(N-1)L$ as $\tau = \frac{(N-1)L}{\upsilon_{g}^{0}}$. $\upsilon_{g}^{0}$ is the group velocity at the frequency, denoted by the circle in Fig.~\ref{DR}(f), at which the group velocity dispersion (GVD)\cite{GVD} is minimum to reduce pulse distortion. Figure \ref{TGT}(b) summarizes the trade-offs among the relevant parameters, i.e., $\tau$, $\Delta l$, bandwidth and maximum transmission for the analyzed case of $N=6$ HRs. The maximum transmission in dB has been calculated by Eq.~(\ref{Trans}). As one can observe, to obtain a large time delay using a fixed number of identical resonators, a very small detuning is needed, i.e., $\Delta l\simeq0$. However, as the detuning is decreased, the bandwidth of the propagating frequencies becomes smaller and the overall losses of the structure become larger. As it is well established in lossy photonic \cite{Pedersen} and phononic \cite{Laude} crystals, losses particularly impact slow wave modes. It is worth noting that modes with group velocities near zero can disappear when losses are considered \cite{Reza}.

\begin{figure*}
\centering 
\includegraphics[width=15cm]{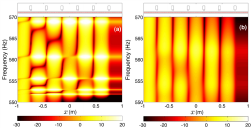}
\caption{(Color online) Numerical sound pressure level, $20\log(|p|)$, obtained using FEM and evaluated for the range of frequencies between 550 Hz and 570 Hz inside the tube loaded with 6 HRs along the line shown in the upper insets (red dashed line) for the lossless (a) and lossy (b) cases with $\Delta=-0.8$ cm.}
\label{6HR}
\end{figure*} 

Numerical simulations using Finite Element Method (FEM) have been performed to highlight the effect of losses in the field distribution inside the waveguide loaded with 6 HRs with $\Delta l =-0.8$ cm (as in Fig.~\ref{TGT}). A plane wave travelling from left to right is considered being the ends of the tube in the numerical domain surrounded by perfectly matched layers in order to numerically approximate the Sommerfeld radiation condition. Figures \ref{6HR}(a) and \ref{6HR}(b) show frequency-position maps of the sound pressure level for the lossless and lossy cases respectively. In the lossless case, the $N-1$ transmitted peaks with the corresponding $N-1$ Fabry-P\'erot resonances inside the finite structure are clearly observed. For these resonant frequencies there is no reflection at the entrance of the tube. However for the lossy case, Fig.~\ref{6HR}(b), the behavior is dramatically modified: the resonances are destroyed and reflections appear at the entrance and thus standing waves are generated. Therefore the system is not any more completely transparent. 
 
\section{Conclusions}
 
In conclusion, we have shown experimentally and theoretically that the presence of losses can drastically influence the slow sound propagation through a periodic locally resonant structure. For the tuned case, i.e the Helmholtz resonance frequency and the Bragg frequency are identical, a super-wide and strongly attenuated band gap appears. For the detuned case, we have shown an experimental group index larger than $20$. We have also investigated in detail the slow sound propagation in a finite lattice of HRs with losses showing that losses reduce and smooth the transmission amplitude of the peaks creating a  broadband of transmitted frequencies in good agreement with the experiments. A trade-off among the relevant parameter (delay time, maximum transmission, bandwidth and detuning) has been presented showing that the near-zero group velocity theoretically predicted disappears due to losses. Finally, using simulated acoustic wave fields into the structure, we have pointed out the presence of reflected waves in the lossy case oppositely to the lossless case. 

We believe that this experimental and theoretical study shows the great importance of losses in acoustic wave propagation through periodic locally resonant structure and contributes to very promising research in the field of acoustic metamaterials, acoustic transmission filters and slow wave applications.

%To summarize the losses reduce the transparency but create a bandwidth of tr

%we have investigated in detail the slow sound propagation in a finite lattice of HRs and
% a narrow transparent peak in- side the band gap, which efficiency is drastically reduced once losses are introduced
 %we investigate in more detail this slow-sound propagation in lossy finite lattice of HRs. In particular, we the slow-sound propagation is characterized using a delay time and we study the trade-offs among this delay and the transmission losses in a finite lattice.
 %one can observe that the effect of losses consists of both reducing the transmission amplitude of the $N-1$ peaks and getting them together creating a broadband of transmitted frequencies in good agreement with the experiments. 
 
 %So, losses reduce the transparency but create a bandwidth of transmitted modes with similar values of transmission.
 %losses particularly impact slow wave modes. It is worth noting that modes with group velocities near zero can disappear when losses are considered.
 
\begin{acknowledgments}
We acknowledge V. Pagneux and A. Maurel for useful discussions. GT acknowledges financial support from FP7-People-2013-CIG grant, Project 618322 ComGranSol. VRG acknowledges financial support from the ``Pays de la Loire'' through the post-doctoral programme.
\end{acknowledgments}


\begin{thebibliography}{99}

\bibitem{Lu} M.-H. Lu, L. Feng and Y.-F. Chen, Materials Today {\bf 12} (12), 34(2009).

\bibitem{Ping} L. Zhengyou, Z. Xixiang, M. Yiwei, Y. Y. Zhu, Y. Zhiyu, C. T. Chan and S. Ping, Science {\bf 289}, 1734 (2000).

\bibitem{Fang} N. Fang, D. J. Xi, J. Y. Xu, M. Ambati, W. Srituravanich, C. Sun and X. Zhang, Nat. Mater. {\bf 5} 452 (2006).

\bibitem{books} P. A. Deymier, {\em Acoustic Metamaterials and Phononic Crystals} (Springer, Heidelberg, 2013); R.~V. Craster and S. Guenneau {\em Acoustic Metamaterials: Negative Refraction, Imaging, Lensing and Cloaking} (Springer, Heidelberg, 2013).

\bibitem{meta} L. Solymar and E. Shamonina, {\it Waves in Metamaterials} (Oxford, University Press, New York, 2009).

\bibitem{Huang04} K.C. Huang, E.Lidorikis, X. Jiang, J.D. Joannopoulos and K. Nelson, Phys. Rev. B, {\bf 69}, 195111, (2004).

\bibitem{Pedersen} J.~G. Pedersen, S. Xiao, N.~A. Mortensen, Phys. Rev. B {\bf 78}, 153101 (2008).

\bibitem{McPhedran} P.~Y. Chen et.al., Phys. Rev. A  {\bf 82}, 053825 (2010).

\bibitem{Moiseyenko11} R.P. Moiseyenko and V. Laude, Phys. Rev. B, {\bf 83}, 064301, (2011).

\bibitem{Hussein09} M.I. Hussein, Phys. Rev. E, {\bf 80}, 212301 (2009).

\bibitem{Laude} R.~P. Moiseyenko, V. Laude, Phys. Rev. B {\bf 83}, 064301 (2011).

\bibitem{Andreassen13} E. Andreassen and J.S. Jensen, J. Sound. Vib. {\bf 135}, 041015, (2013).

\bibitem{Zwikker} C.~Zwikker and C.~W. Kosten, \newblock {\em Sound absorbing materials}, (Elsevier Publishing Company, Inc., Amsterdam, 1949).

\bibitem{Xiao} Y. Xiao, B. ~R. Mace, J. Wen, X. Wen, Phys. Lett. A {\bf 375}, 1485 (2011).

\bibitem{Paris} N. Kaina, M. Fink, G. Lerosey, submitted.

\bibitem{Page} C. Cro{\"e}nne, E. J. ~S. Lee, H. Hu, J. ~H. Page, AIP Advances {\bf 1}, 041401 (2011). 

\bibitem{Wang} X. Weng, C. Mak, J. Acoust. Soc. Am. {\bf 131}(2), 1172 (2011).

\bibitem{Sugimoto} N. Sugimoto and T. Horioka, J. Acoust. Soc. Am. {\bf 97}(3), 1446 (1995).

\bibitem{Seo} S.-H. Seo, Y.-H. Kim, J. Acoust. Soc. Am. {\bf 118}(4), 2332 (2005).

\bibitem{Boudouti} E.H. El Boudouti et. al., J. Phys.: Condens. Matter {\bf 20}, 255212 (2008).

\bibitem{PRERichoux} O. Richoux, V. Tournat, T. Le Van Suu, Phys. Rev. E {\bf 75}, 026615 (2007).

\bibitem{Bradley} C.~E. Bradley, J. Acoust. Soc. Am. {\bf 96}(3), 1844 (1994).

\bibitem{EPLOlivier} O. Richoux, V. Pagneux, Europhys. Lett., {\bf 59}(1), 34 (2002).

\bibitem{Pierce} A.~D. Pierce, {\em Acoustics : An introduction to its physical principles and applications}, (Mac Graw Hill, 1981).

\bibitem{ImpedanceSensor} C.~A. Macaluso, J.-P. Dalmont, J. Acoust. Soc. Am. {\bf 129}(1), 404 (2011).

\bibitem{Lissek} F. Bongard, H. Lissek, and J. ~R. Mosig, Phys. Rev. B {\bf 82}, 094306 (2010).

\bibitem{Caloz}  C. Caloz and T. Itoh {\em Electromagnetic Metamaterials: Transmission Line Theory and Microwave Applications}, (Wiley-Interscience and IEEE Press, Hoboken, NJ, 2006).  

\bibitem{slowsound} A. Santill\'an and S. I. Bozhevolnyi, Phys. Rev. B {\bf 84}, 064304 (2011).

\bibitem{GVD} Once the group velocity depends on frequency in a medium, one can introduce the Group Velocity Dispersion (GVD), GVD$=\partial^2k/\partial\omega^2$. This parameter is for example used in optics for the analysis of the dispersive temporal broadening or compression of pulses.

\bibitem{Reza} A. Reza, M.~M. Dignam, and S. Hughes, Nature (London) {\bf 455}, E10 (2008).






%
%\bibitem{Yariv} J. B. Khurgin and R. S. Tucker, {\em Slow Light : Science and Applications}, (CRC Press
%Taylor and Francis Group, 2009).
%
%\bibitem{Yariv1} Y. Xu, Y. Li, R. K. Lee, and A. Yariv, Phys. Rev. E. {\bf 62}, 7389 (2000).
%
%\bibitem{Boller} K. J. Boller, A. Imanoglu, and S. E. Harris, Phys. Rev. Lett. {\bf 66}, 2593 (1991).
%
%\bibitem{Boris} B. Luk'yanchuk et.al., Nature Mat. {\bf 9}, 707 (2010).
%
%\bibitem{Fano} A. E. Miroshnichenko, S. Flach, and Y. Kivshar, Rev. Mod. Phys. {\bf 82}, 2257 (2010).
%
%\bibitem{optics1} Q. Xu et.al., Phys. Rev. Lett. {\bf 96}, 123901 (2006).
%
%\bibitem{optics2} X. Yang et.al., Phys. Rev. Lett. {\bf 102}, 173902 (2009).
%
%\bibitem{acoustics} A. Santill\'{a}n and S. I. Bozhevolnyi, Phys. Rev B, {\bf 84}, 064304 (2011); F. Liu et. al., Phys. Rev. E. {\bf 82}, 026601 (2010).
%
%\bibitem{Laude} R.~P. Moiseyenko, V. Laude, Phys. Rev. B {\bf 83}, 064301 (2011).
%
%\bibitem{Tsakmakidis} K.L. Tsakmakidis, A.D. Boardman and O. Hess. Nature (London), {\bf 455}, E10, (2008)
%
%\bibitem{com} These effects are minimized as the number of the unit cells is increased, see also section 3.2.8 in \cite{Caloz}. However, in our case, as the number of the HRs is increased, due to the high attenuation around the resonance frequency, the level of the received signal is too low and thus the phase information is lost. An experimental set-up to overcome these difficulties is in progress. 
%
%\bibitem{Soukoulis} P. Marko\u{s} and C.~M.Soukoulis {\em Wave Propagation}, (Princeton University Press, Princeton, NJ, 2008). 

\end{thebibliography}
\end{document}